\documentclass[12pt]{article}
\usepackage{graphics}
\usepackage{epsfig}
\usepackage{graphicx}
\usepackage{graphicx}
\newcommand{\ve}{\varepsilon}
\newcommand{\pl}{\partial}
\renewcommand\thesubsubsection{\arabic{subsubsection}}

\begin{document}

\begin{center}
{\large\bf Emission tomography of laser induced plasmas 
with large acceptance angle
apertures} 

\vskip 1cm
{\bf S.~V.~Shabanov${}^{1,2}$ and 
I.~B.~Gornushkin${}^2$}

\vskip 1cm
${}^1$ {\it Department of Mathematics, University 
of Florida, Gainesville, FL 32611, USA}
%${}^*$Corresponding author:\ \ shabanov@math.ufl.edu

${}^2$ {\it BAM Federal Institute for Materials Research and Testing, Richard-Willst\"atter-Strasse 11, 12489 Berlin, Germany}

\end{center}

\begin{abstract} 
It is proposed to use apertures with large acceptance angles
to reduce the integration time when studying the emissivity
of laser induced
plasmas by means of the Abel inversion method. 
The spatial resolution lost due to contributions of 
angled lines of sight to
the intensity 
data collected along the plasma plume diameter is restored by a special
numerical data processing. The procedure is meant for
the laser induced plasma 
diagnostics and tomography when the integration time
needed to achieve a reasonable signal to noise ratio 
exceeds  a characteristic time scale of the plasma state variations
 which is short especially at early stages
of the plasma evolution.

\end{abstract}

\subsubsection{LIBS plasma tomography experiments}

\begin{figure}
 \centering
\includegraphics[height=5cm,width=12cm]{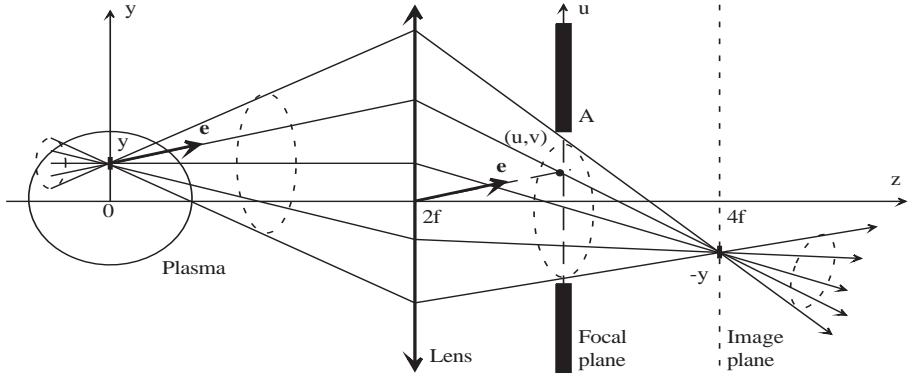}
\caption{\small A typical experimental setup for LIBS plasma
measurements. A plasma plume is created by a laser ablation at 
the distance of two focal length from a lens. The light intensity 
is measured in the image plane. An aperture is placed in the lens
focal plane. It restricts the acceptance cone at each point where
the intensity is measured (provided the acceptance cone of the light
collecting device is wider than that of the aperture). Photons
propagating along any particular line of sight parallel to ${\bf e}=
{\bf e}_{\theta,\varphi}$ that lies within the solid angle 
of the acceptance cone and goes though the point $(0,y,0)$ in 
the plasma plume are collected at the point $(0,-y,4f)$.
The measurements give the intensity $I_\Omega(y)$ as a function
of $y$. For large acceptance cones, $I_\Omega(y)$ cannot be viewed
as the intensity in parallel lines of sight and, hence, cannot be 
directly used in the Abel inversion to reconstruct the plasma emissivity.
}
\label{fig1}
\end{figure}

Laser-induced breakdown spectroscopy (LIBS) is widely used
for quantitative chemical analysis of 
an elemental content of various materials \cite{a2,a1}. A typical 
experimental setup (see, e.g., \cite{5,6,7}) is depicted 
in Fig.~\ref{fig1}. A plasma plume is created by a laser
ablation at the distance of two focal lengths from a lens.
The lens projects the plasma plume image onto the image plane 
where a detecting device is placed (e.g., a spectrometer slit).
The plasma plume expands, cools, and radiates. 
All photons propagating along a particular line of sight
parallel to a vector ${\bf e}$ through a point at the distance $y$
from the plasma plume symmetry axis can be collected at the point
$(0,-y, 4f)$, e.g., by a spectrometer slit. The intensity $I_\Omega(y)$
measured at the point $(0,-y,4f)$ depends on the solid angle $\Omega$
in which the light is collected. If $\Omega$ is small, then 
the intensity $I_\Omega(y)$ can be approximately viewed as the 
intensity $I(y)$ in (nearly) parallel lines of sight. In this
approximation, the measured intensity is related to the plasma
emissivity by the Abel integral equation that can be solved
to recover the emissivity \cite{1}. For this reason, spectrometers
used in these measurements must have a high $f-$number in order
to restrict the solid angle of the acceptance cone. 
The latter 
implies, in particular, a small amplitude of the measured signal per 
unit time. In turn, a long integration time is required to achieve
a reasonable signal to noise ratio. If the plasma plume remained
stationary during the integration time, the reconstructed
emissivity (as well as the plasma temperature distribution and 
local densities of electrons, ions, and atoms) would have 
corresponded to the true
(instantaneous) plasma parameters. It appears, however,
the plasma plume cannot be viewed as stationary during a typical
integration time of $1 \mu s$ or higher. 
This is especially true for early
stages of the plasma plume evolution \cite{3,4}.     
 
Here the following possibilities are discussed to reduce the 
integration time. If the plasma plume has the {\it spherical symmetry}
(or, at least it can be approximated as such), then
it is shown that any device with an arbitrary 
large acceptance cone
can be used to collect the data. In particular, spectrometers
with low $f-$numbers can be used. The loss of the spatial resolution
due to contributions of angled lines of sight into the data function
$I_\Omega(y)$ can be eliminated by a proposed numerical data processing
such that the intensity $I(y)$ in parallel lines of sight is accurately
recovered and can be used in the subsequent Abel inversion to obtain
the plasma emissivity. The data processing is particularly simple
if the device acceptance cone is slightly restricted by a 
large aperture of a special shape (see Section 3) 
placed in the lens focal plane
(as depicted in Fig.~\ref{fig1}).
But with minor modifications
the reconstruction algorithm is shown to 
work for any acceptance cone of the 
measuring device (or any aperture).

If the plasma plume has the {\it axial symmetry}, then a measuring 
device may also have an arbitrary large acceptance cone, but 
the lines of sights that are not perpendicular to the plasma
plume symmetry axis must be blocked by an aperture in the focal
plane whose dimension
along the symmetry axis is sufficiently small. In other words,
only lines of sight that lie within a thin slice of the plasma plume
(normal to the symmetry axis) contribute to the measured intensity.
This is similar to the conventional experimental setup (e.g., a spectrometer with a high $f-$number). The difference
is that photons propagating along 
all angled lines of sight within the plasma plume slice 
can be collected to amplify the measured signal. A simple numerical data
processing algorithm is proposed to accurately recover the intensity 
in parallel lines of sight.

So, a rational behind the use of larger apertures is a reduction 
of the integration time. This is especially important when studying
such a dynamical object as laser-induced plasmas. The plasma 
diagnostics and tomography based on the Abel inversion becomes
inaccurate when the integration times exceeds 
a characteristic time scale of the plasma state variations
 which is short especially at early stages
of the plasma evolution (e.g., a  few $ns$ after the ablation process
is over). The method proposed below would allow for a sufficient
signal to noise ratio with integration times in the range of $10$'s to
$100$'s $ns$ at the price of loosing a spectral resolution.
It also allows for the use of spectrometers with low $f-$numbers,
which reduces the integration time, but not as much as in 
the case when the spectral resolution  
is not needed. 

\subsubsection {A generalized Abel equation for apertures with a 
large acceptance angle} 
Suppose that a plasma plume is created on the optical
axis of a lens that is at a distance of two focal lengths
from the plume (see Fig.~\ref{fig1}). 
Then an image of the plasma plume is 
created in the plane normal to the optical axis that is 
at the distance of two focal lengths but on the other side 
of the lens. Let the plasma emissivity be given as 
a function of the position vector ${\bf r}$, $\ve=\ve({\bf r})$.
The question to be studied is the amount of plasma radiation 
that can be collected at a point in the image plane. 
The coordinate system is set so that the optical axis coincides
with the $z-$axis. Consider an infinitesimal area element
$d\sigma=dxdz$ at the point $(0,y,0)$ (the origin is inside the plasma
plume) normal to the $z-$ axis. 
Let ${\bf e}_{\theta,\varphi}=
(\cos\varphi\sin\theta,\,\sin\varphi\sin\theta,\,\cos\theta)$
be the unit vector that defines the direction of 
a line of sight through the point $(0,y,0)$. The geometrical
meaning of the angles $\theta$ and $\varphi$ is transparent.
The line has the angle $\theta$ with the optical axis. Variations
of $\varphi$ correspond to lines that lie on the cone of 
the angle $\theta$ whose 
the axis passes through $(0,y,0)$ and is parallel to the optical
axis. By letting $\theta$ and $\varphi$ range over the 
intervals $[0,\pi/2]$ and $[0,2\pi]$, respectively,
all lines of sight through the point $(0,y,0)$ in the plasma
plume are parameterized. Without any restriction on the lens
diameter, all such lines intersect at the point $(0,-y,4f)$
in the image plane, where $f$ is the lens focal length.

A lens has a finite size and yet it is subject to optical 
aberration effects. For this reason a diaphragm should be placed
in the lens focal plane (see Fig.~\ref{fig1}). The diaphragm 
aperture restricts the "emission" solid angle within which
the light can be collected at the point $(0,-y,4f)$.
As noted before, spectrometers with high $f-$numbers
are used to collect the light (the spectrometer slit 
is placed in the image plane parallel to the $y-$ axis). 
For a high $f-$number, the spectrometer acceptance solid angle
is smaller than the "emission" solid angle. So, not all the 
plasma radiation that arrives at the observation point is 
collected. If the acceptance solid angle is small, then 
the measured intensity is proportional to the intensity 
in (nearly) parallel lines of sight going through the plasma
plume diameter (see below) that is used in the Abel inversion method. 

Let us put aside the question about 
how exactly the light is collected at the point $(0,-y,4f)$
and assume that all the light coming through the aperture
can somehow be collected. Alternatively, one can think of 
any measuring device that has a large acceptance cone.
As is clear from Fig.~\ref{fig1}, any acceptance 
cone is equivalent to some aperture placed in the lens focal
plane. So, measuring devices and apertures restricting 
the "emission" cone can be discussed on the same footing.

The energy collected per unit time,
unit frequency, and unit area 
that was emitted into a solid angle $d\Omega$ 
in the direction of ${\bf e}_{\theta,\varphi}$
reads \cite{chanrasekar}
\begin{equation}
\label{1}
I_\Omega(y) = \int_\Omega d\Omega \cos\theta \int_{-\infty}^{\infty}
ds\, \ve\Bigl({\bf r}_{\theta,\varphi}(s,y)\Bigr)\equiv
\int_\Omega d\Omega \cos\theta\ I(\theta,\varphi,y)
\,, 
\end{equation}
where the dependence on the frequency and time is not explicitly
shown (it is not relevant for what follows),
${\bf r}_{\theta,\varphi}(s,y) = (0,y,0) + 
{\bf e}_{\theta,\varphi} s$ is the parametric equation of
the line of sight through the point $(0,y,0)$ and parallel
to the unit vector ${\bf e}_{\theta,\varphi}$, $s$ is 
the arclength parameter,
$d\Omega = 
d\theta d\varphi \sin\theta$, and the range for the angles
$\theta$ and $\varphi$ is determined by the shape of 
the aperture in the focal plane. The function $I(\theta,\varphi,y)$
is the light intensity along the line of sight parallel
to ${\bf e}_{\theta,\varphi}$ through the point $(0,y,0)$ in 
the plasma plume. To find the dependence of $I_\Omega$ on
the aperture geometry, the integration over 
the solid angle is transformed to a double integral 
over the aperture by a change of variables. Let $(u,v)$ be 
rectangular coordinates in the focal plane such that the origin
lies on the lens optical axis and the $u-$axis is parallel to the 
$y-$axis (see Fig.~\ref{fig1}). 
If $dA=dudv$ is the area element at the aperture
point $(u,v)$, then 
$$
d\Omega = \frac {dA \cos\theta}{f^2+u^2+v^2}
$$
and the relation between the angles and rectangular coordinates
in the focal plane is given by
\begin{equation}
\label{2}
v=f\tan\theta \cos\varphi\,,\quad u=f\tan\theta\sin\varphi\,.
\end{equation}
Equation (\ref{1}) can be written in the form
\begin{equation}
\label{3}
I_\Omega(y) = \int_A dA\ \frac{f^2}{(f^2+u^2+v^2)^2}\, 
I(\theta,\varphi,y)\,,
\end{equation}
where the angles $\theta$ and $\varphi$ in the arguments of $I$
are expressed via $u$ and $v$ by inverting relations (\ref{2}).
The integration in (\ref{3}) is carried out over 
the aperture area denoted by $A$.

Consider first the case of a spherically symmetric plasma plume
(the case of the axial symmetry is discussed in Section 4), i.e.,
 $\ve ({\bf r})=\ve(r)$ where 
$r=|{\bf r}|$. The distance  $r_{\theta,\varphi}(y)$ 
between the line ${\bf r}=
{\bf r}_{\theta,\varphi}(s,y)$ and the origin (the plasma plume
center) is given by the magnitude of the vector that is 
the cross product $(0,y,0)\times {\bf e}_{\theta,\varphi}$.
A simple calculation shows that $r_{\theta,\varphi}(y) = 
|(0,y,0)\times {\bf e}_{\theta,\varphi}| = 
y(1-\sin^2\theta\sin^2\varphi)^{1/2}$ for $y\geq 0$.
If $I(0,0,y)=I(y)$ is the intensity along the line of sight
parallel to the optical axis, then it follows from the assumption
of the spherical symmetry of the plasma plume that
\begin{equation}
\label{3a}
I(\theta, \varphi, y) = I(r_{\theta,\varphi}(y))\,,\quad\quad
r_{\theta,\varphi}(y)=y\sqrt{1-\sin^2\theta\sin^2\varphi}
\end{equation}
because the intensity along any line of sight through 
the point $(0,y,0)$ depends only on the distance of 
that line to the plasma plume center. By making use 
of relations (\ref{2}) to express $r_{\theta,\varphi}$
as a function of $u$ and $v$, Eq. (\ref{3}) can be written 
in the form
\begin{equation}
\label{4}
I_\Omega(y) = \int_A dA\ \frac{f^2}{(f^2+u^2+v^2)^2}\, 
I\Bigl(y\mu(u,v)\Bigr)\,,\quad\quad
\mu(u,v) = \sqrt{\frac{f^2+v^2}{f^2+u^2+v^2}}\,.
\end{equation}
If $A$ is an infinitesimal aperture of the area $\Delta A$ 
centered on the optical
axis, then $I_\Omega(y) \approx  \Delta \Omega I(y)$ where 
$\Delta \Omega = \Delta A/f^2$.  In this case,
the data function can be used as the intensity 
along parallel lines of sight 
$I(y) \approx I_\Omega(y)/\Delta\Omega$ that is related 
to the plasma emissivity $\ve(r)$ 
by the 
the Abel integral equation
\begin{equation}
\label{5}
I(y) =  \int_{-\infty}^\infty
ds\,\varepsilon\Bigl(\sqrt{y^2+s^2}\Bigr)=2
\int_y^\infty \frac {dr\, r\, \ve(r)}{\sqrt{r^2-y^2}}\,.
\end{equation}
It can be solved for $\ve(r)$.
The procedure is known as the Abel inversion \cite{1}:
$$
\ve(r) = 
 -\frac 1\pi \int_r^\infty dy \
\frac{I'(y)}
{\sqrt{y^2 - r^2}}= \int_0^\infty dk\,kJ_0(kr)\tilde{I}(k)\,,\quad
\quad \tilde{I}(k) =\int_{-\infty}^\infty dy e^{iky}I(y)\,,
$$ 
where $\tilde{I}(k)$ is the Fourier transform of $I(y)$ and
$J_\nu$ is the Bessel function of index $\nu=0$. 
Substituting (\ref{5}) into the right side of Eq. (\ref{4}),
an analog of the Abel integral equation (\ref{5}) is obtained
for an aperture with a finite acceptance angle. 
If Eq. (\ref{4}) can be solved for $I(y)$ for a given $I_\Omega(y)$,
then a solution to the generalized Abel equation 
is found by means of the Abel inversion for $I(y)$.
In Sections 5 a numerical algorithm is proposed for
solving Eq. (\ref{4}) in the case of an aperture of a special
shape. The case of an aperture of a general shape is presented
in Section 6. A numerical example of is given in Section 7. 

\subsubsection {The hyperbolic aperture}

It turns out that there is a particular shape of the aperture
for which Eq. (\ref{4}) is substantially simplified and becomes
convenient for a qualitative analysis of the relation 
between the intensity in parallel lines of sight $I(y)$ and 
the data function 
$I_\Omega(y)$ for large apertures.

\begin{figure}
 \centering
\includegraphics[height=5cm,width=14cm]{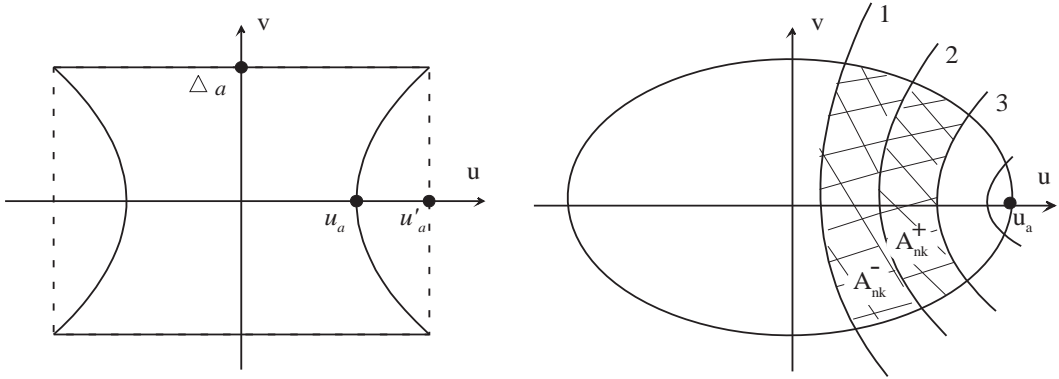}
\caption{\small {\sf Left panel:} A hyperbolic aperture
that allows for a factorization in the sum over angled lines
of sight. It is bounded by the lines $v=\pm \Delta_a$ and
hyperbolas $u =\pm \tan\theta_a\sqrt{v^2+f^2}$.
Here $u_a = f\tan\theta_a = f\sqrt{1-\mu_a^2}/\mu_a$
and $u_a' = \tan\theta_a \sqrt{f^2 + \Delta_a^2}$. For 
$\Delta_a\sim f$, the aperture cannot be approximated by a 
rectangle. A rectangular approximation is accurate if 
$\Delta_a\ll f$. It can be used for studies of axially symmetric
plasmas as explained in the text.
{\sf Right panel:} An illustration to the reconstruction algorithm
for a general aperture. Curves 1, 2, and 3 are the level curves (hyperbolas)
$\mu(u,v)=\mu_{nk}^-$, $\mu(u,v)=\mu_{nk}^+$, and 
$\mu(u,v)=\mu_{nk+1}^-$, respectively, that are used to partition
the region occupied by the aperture. In each partition region
$A_{nk}^\pm$, the function $I_n(\mu)$ is approximated by a linear
function as is explained in the text. Here $\mu_a$ is the minimal
value of the function $\mu(u,v)$. }
\label{fig2}
\end{figure}

Let $A$ be symmetric relative to the reflection $u\rightarrow -u$
and let $A_+$ be the half of $A$ that lies in the half-plane 
$u\geq 0$. Then $\int_A = 2\int_{A_+}$ in the right side of 
Eq. (\ref{4}). Consider the change of variables in the double 
integral (\ref{4}): $(u,v)\rightarrow (\mu,v)$ where 
the new variable $\mu =\mu(u,v)$ is defined by the second relation
in (\ref{4}). Then $dA=dudv = dvd\mu J$ where the Jacobian
is $J=\pl u/\pl \mu = (\pl \mu/\pl u)^{-1}$. The coordinate 
line $\mu=const$ is a hyperbola in the $(u,v)-$plane:
\begin{equation}
\label{tail}
u= \frac{\sqrt{1-\mu^2}}{\mu}\sqrt{v^2 +f^2}\,.
\end{equation}
It intersects the $u-$axis at $u=f\sqrt{1-\mu^2}/\mu$ and has 
slant asymptotes $u=|v|\sqrt{1-\mu^2}/\mu$ at large $|v|$. 
In particular, the line $u=0$ (or the $v-$axis) is the limiting
hyperbola as $\mu \rightarrow 1$. Note that the $u-$intercept
of the hyperbola (\ref{tail}) tends to zero, 
while the slope of the slant
asymptotes becomes infinite, i.e., the hyperbola tends to the 
vertical straight line through $u=0$. Let $A_+$ be bounded 
by horizontal lines $v=\pm \Delta_a$, the vertical line $u=0$,
and by the hyperbola (\ref{tail}) with $\mu=\mu_a <1$ as is shown
in the left panel of Fig.~\ref{fig2}.
Then after 
the change of integration variables, the range of new 
variables is the rectangle $-\Delta_a \leq v\leq \Delta_a$
and $\mu_a\leq \mu\leq 1$. A straightforward
calculation shows that the double integral (\ref{4}) is factorized
in the new variables thanks to the "hyperbolic" shape of 
the aperture:
\begin{equation}\label{7}
I_\Omega (y) = 4\int_0^{\Delta_a} \frac{dv}{\sqrt{f^2+v^2}}
\int_{\mu_a}^1 \frac{d\mu\, \mu^2}{\sqrt{1-\mu^2}}\ I(\mu y)=
\frac{4\Delta_a}{\sqrt{f^2+\Delta_a^2}}
\int_{\mu_a}^1 \frac{d\mu\, \mu^2}{\sqrt{1-\mu^2}}\ I(\mu y)\,.
\end{equation}
Put $\mu=\cos\theta$ so that $u=f\tan\theta_a\equiv u_a$ 
defines the $u-$intercept
of the hyperbola (\ref{tail}) with $\mu=\mu_a$
that bounds the aperture area $A_+$. 
Put also $\tan\varphi_a =\Delta_a/f$.
Then 
\begin{equation}
\label{8}
I_\Omega(y) = 4\sin\varphi_a \int_0^{\theta_a} d\theta \cos^2\theta\
I(y\cos\theta)\,.
\end{equation}
This equation is simple enough to estimate the signal amplification
due to an aperture with a large acceptance angle and compare it with 
 signal measurements in reported LIBS plasma tomography
experiments \cite{6,7}.

In the conventional LIBS plasma tomography experiments,
the slit of a spectrometer is placed into the image plane
parallel to $y-$axis. However, not all the light that enters
into the spectrometer slit at
the point $(0,-y,4f)$ is collected but rather its only portion
that illuminates a spectrometer mirror (that further directs 
the light onto the spectral grating). 
If the aperture in the focal plane is not too small or not 
present at all,
this portion is determined only
by the $f-$number of the spectrometer so that the acceptance 
solid angle in Eq. (\ref{1}) is defined by $0\leq \varphi\leq 2\pi$
and $0\leq \theta\leq \theta_s$ where $\tan\theta_s=1/(2f_s)$
and $f_s$ is the spectrometer $f-$number. 
Spectrometers in 
the reported LIBS plasma tomography experiments typically have 
relatively large $f-$numbers, $f_s\sim 5$. So, the acceptance 
angle is small, $\tan\theta_s \approx \theta_s \sim 10^{-1}$.
It is then sufficient to calculate the integral (\ref{1}) in the 
leading order of $\theta_s^2$, while neglecting terms of order
$\theta^4_s$. In this approximation $I(\theta,\varphi,y)=
I(y) + O(\theta^2)$ (cf. (\ref{3a})) and the evaluation of the 
integral (\ref{1}) yields:
$$
I_\Omega (y) = \pi\theta_s^2 I(y) + O(\theta^4_s)\,.
$$
The smallness of the factor $\pi\theta_s^2$ in the rate of the energy
flux per unit time
is exactly the reason
to have a long integration time to get a sufficient signal to noise
ratio, when determining $I(y)$ used in the Abel inversion.
Suppose now that all the light that comes to the point $(0,-y,4f)$ 
through the hyperbolic aperture 
is collected. Assuming, for sake of simplicity, 
$I(y\cos\theta)$ to be 
a constant $I$,  the evaluation of the integral in (\ref{8})
yields:
$$
I_\Omega \approx 4\sin\varphi_a (2\theta_a -\sin(2\theta_a))\,I\,.
$$
So the signal amplification (as compared to the spectrometer case)
can be estimated by a factor of 
$4\sin\varphi_a (2\theta_a -\sin(2\theta_a))/(\pi\theta_s^2) \sim 50$
if $\theta_s\sim 10^{-1}$ and
$\varphi_a\sim\theta_a\sim \pi/4$ (i.e., $\Delta_a\sim u_a\sim f$).

To increase the angles $\varphi_a$ and $\theta_a$, spectrometers
with small $f-$numbers can be used. An effective aperture
corresponding to this case is a circular disk of the radius
$f/(2f_s)$. If the hyperbolic aperture is fit into this disk,
then Eq. (\ref{8}) applies directly, and $I(y)$ is recovered 
by the numerical data processing proposed in Section 5. Otherwise,
a circular aperture should be treated by the algorithm of 
Section 6.

If the spectral resolution is not needed, then a gated ICCD camera
can be placed in the image plane. The pixel row of the camera
is set along the plasma image diameter. The light collected by 
each pixel provides the spatially resolved data $I_\Omega(y)$.
If the aperture has the hyperbolic shape, Eq. (\ref{8}) applies
directly to recover $I(y)$. Otherwise, the algorithm of Section 6
must be used. To study the plasma emissivity in a narrow
frequency window of interest, a narrow band (or interference) filter
can be positioned in front of the detector (the pixel array).
This method can be particularly useful for diagnostics 
of laser induced plasmas shortly after the ablation process ($\sim\,
10$'s $ns$). At this stage 
the plasma exhibits a fast dynamics and emits mostly
a continuum spectrum radiation.

\subsubsection{Axially symmetric plasma plumes}

For axially symmetric plasma plumes, photons propagating 
along lines of sight that are not perpendicular to the 
plume symmetry axis must be prevented from being collected
by a detector so that only a plasma plume slice normal to the 
symmetry axis is visible to a measuring device.
For spectrometers with high $f-$ numbers,
this is achieved automatically due to their narrow acceptance cones.
But a narrow acceptance cone also eliminates most of lines of sight
in the visible slice itself that are angled to the optical axis.
If $R$ is a plasma plume radius, then the lines of 
sight through $(0,y,0)$ in the acceptance cone of the angle
$\theta_s$ can go through plasma points
separated 
 at most by the distance $2R\tan\theta_s$.
This determines the spatial resolution along the $y-$axis
(as well as along the symmetry axis). 
For $\theta_s \sim 10^{-1}$, the resolution is such that the data can be collected at about
10 positions across the plasma plume diameter without a significant 
overlap of the plasma regions from which the light is collected. 
So, when 
spectrometers with lower $f-$numbers or pixel rows of 
an ICCD camera (as suggested above) are used to  increase the signal,
the spatial resolution become even worse along both the directions,
the $y-$ and  symmetry axes.

The following compromise can be made.
The spatial resolution in the direction 
parallel to the symmetry axis
is achieved by placing a
slit-like aperture extended along the $y-$axis in the focal plane.
If a slit of height $2\Delta_a$ is positioned in the focal plane
($-\Delta_a\leq v\leq \Delta_a$ in the above notations),
then only photons traveling along 
the lines of sight that are in 
the plasma "slice" of height $2R\tan\varphi_a=2R\Delta_a/f$
can reach the image plane.
The shape of a sufficiently narrow rectangular aperture 
can well be 
approximated by the hyperbolic aperture with small $\Delta_a$
because the hyperbola (\ref{tail}) has the vertical 
tangent line at the intercept point $u=u_a$ (see the caption 
of Fig.~\ref{fig2}). Thus, 
Eq. (\ref{7}) or (\ref{8}) remains applicable in this case, but 
the angle $\theta_a$ is determined by 
the measuring device
$f-$number, i.e., $\tan\theta_a = 1/(2f_s)$ (the approximation
of a small $\theta_s$ is invalid for small $f_s$), and 
$\varphi_a$ is small enough to provide for the spatial
resolution along the plasma plume symmetry axis. 

Two-dimensional detectors used in LIBS experiments \cite{6,7} in
combination with a spectrometer typically have about
a few hundred pixels for the spatial resolution along 
the spectrometer slit, while only 10 to 20 spatially resolved
data points are taken (each data point corresponds to an average
intensity over a cluster of pixels). 
The algorithm provided below allows for an accurate 
reconstruction of the spatial resolution of just 
a single pixel size. So, it may not be deprived of sense
to apply the proposed algorithm to improve the spatial resolution
even in spectrometers with higher $f-$numbers. This, however,
depends very much on the gradients of thermodynamic parameters
of the observed plasma. For high gradients, this is definitely
the case.

\subsubsection{The reconstruction algorithm}

Let the light be collected by a linear array of pixel detectors
placed along the $y-$axis in the image plane. Each pixel has
a size $\Delta$. The pixel centers are located at $y=y_n=n\Delta$,
$n=0,1,...,N$ (by the symmetry of the observed plasma, it is
sufficient to reconstruct $I(y)$ (the intensity in parallel lines
of sight) only for non-negative $y$, $I(-y)=I(y)$). In what 
follows the factor $4\sin\varphi_a$ in (\ref{8}) will be omitted.
The energy of all photons collected by a pixel $n$ per unit time
and unit frequency is given by 
\begin{eqnarray}
\label{a1}
I^\Omega_n &=& \int_{y_{n-1/2}}^{y_{n+1/2}} dy I_\Omega(y) =
\int_{\mu_a}^1 \frac{d\mu\,\mu^2}{\sqrt{1-\mu^2}}\, 
\int_{y_{n-1/2}}^{y_{n+1/2}} dy I(\mu y) = 
\int_{\mu_a}^1 \frac{d\mu\,\mu}{\sqrt{1-\mu^2}}\, I_n(\mu)\quad\\
\label{a2}
I_n(\mu)&=& \int_{\mu y_{n-1/2}}^{\mu y_{n+1/2}} du\ I(u)\,,
\end{eqnarray}  
where $y_{n\pm 1/2}= \Delta (n\pm 1/2)$ are the boundary points
of the $n$th pixel, and the change of the integration variable
$u=\mu y$ has been carried out. If only the photons propagating 
along parallel lines of 
sight were registered by the $n$th pixel, then the data would have 
consisted of values
\begin{equation}
\label{a3}
I_n = \int_{y_{n-1/2}}^{y_{n+1/2}} du \ I(u) = I_n(1)\,.
\end{equation}
Exactly these values must be used in the Abel inversion to
reconstruct the emissivity. The idea is to find a suitable 
interpolation for the functions $I_n(\mu)$ using the "sampling"
points $I_n$. Then the integral with respect to $\mu$ in (\ref{a1})
can be evaluated thus turning Eq. (\ref{a1}) into a system
of algebraic equations for $I_n$ that can be solved for a given
data set $I_n^\Omega$.

Let $P_n(\mu)$ denote the interval $[\mu y_{n-1/2},\mu y_{n+1/2}]$
so that $P_n(1)=P_n$ is the interval occupied by the $n$th pixel.
Depending on the values of $\mu$ and $n$, the interval
$P_n(\mu)$ may have the following positions relative to 
the intervals occupied by physical pixels:
\begin{equation}
\label{a4}
P_n(\mu)\ {\rm overlaps\ with\ both}\ P_{n-k+1}\ 
{\rm and}\ P_{n-k},\quad {\rm or}\quad P_n(\mu)\ {\rm is\ 
contained\ in}\ P_{n-k}
\end{equation}
for $k=1,2,..., N_n$ where $N_n = [(1-\mu_a)n] +1$ and
$[x]$ denotes the largest integer that does not exceed $x$.
The integer $N_n$ is defined by the condition 
$y_{n-N_n}< \mu_a y_{n}\leq  y_{n-N_n+1}$. For a fixed value
of $n$ it is not difficult to find the ranges of $\mu$ 
in both the cases (\ref{a4}):
\begin{eqnarray}
\label{a5}
\mu_{nk}^+ \leq &\mu& \leq \mu_{nk}^-\,,\quad {\rm or}\quad
\mu_{nk+1}^- \leq \mu \leq \mu_{nk}^+\,,\\
\label{a6}
\mu_{nk}^\pm &=& \frac{n-k+1/2}{n\pm 1/2}\,,\quad
k=1,2,...,N_n\,,\quad \mu^-_{nN_n+1} =\mu_a\,.
\end{eqnarray}
For example, for $k=1$, the first case in (\ref{a4}) holds
as long as $\mu y_{n+1/2}$ remains in $P_n$, i.e.,
$\mu y_{n+1/2} \geq y_{n-1+1/2}$ or $\mu \geq \mu_{n1}^+$.
As $\mu$ becomes less than $\mu_{n1}^+$, the whole interval $P_n(\mu)$
remains in $P_{n-1}$ until the left boundary
of $P_n(\mu)$  crosses the left boundary of $P_{n-1}$,
i.e.,  $\mu y_{n-1/2} \geq y_{n-2+1/2}$ or $\mu \geq\mu_{n2}^{-}$,
and so on. A special consideration is required for the case
$k=N_n$ because, depending on the value of $\mu_a$, either both 
the cases (\ref{a4}),  or just the first one, or none of them 
may occur. This issue will be 
clarified shortly. For time being, the values $\mu_{nk}^\pm$ are 
defined by (\ref{a6}), i.e., $\mu_a$ is assumed
not to exceed $\mu_{nN_n^+}$ (both the cases (\ref{a4}) are possible
for $k=N_n$).

Consider two integrals $\int_{a}^bdx\, f(x)$ and
$\int_{a'}^{b'}dx\, f(x)$ for some $f(x)$ where
$a\leq a'<b'\leq b$. How can the second integral be 
approximated by the first one? The following approximation 
is proposed:
\begin{equation}
\label{a7}
\int_{a'}^{b'}dx\, f(x) \approx \frac{b'-a'}{b-a}\, 
\int_a^b dx\, f(x)\,.
\end{equation}
The accuracy of this approximation is discussed in Appendix.
For a small interval length $b-a$ and smooth $f$, this is 
a good approximation as is shown in Appendix. If 
$\mu$ is such that the interval $P_n(\mu)$ is contained
in $P_{n-k}$, then the approximation (\ref{a7}) applies
directly:
\begin{equation}
\label{a8}
I_n(\mu) = \frac{\mu\Delta}{\Delta}\ I_{n-k}=
\mu\, I_{n-k}\,,\quad \mu_{nk+1}^-\leq \mu \leq \mu_{nk}^+\,.
\end{equation}
If $\mu$ is such that $P_n(\mu)$ overlaps with $P_{n-k}$
and $P_{n-k+1}$, then $I_n(\mu)$ is approximated by 
a linear combination of $I_{n-k+1}$ and $I_{n-k}$ with 
the coefficients being the lengths of fractions of $P_n(\mu)$
in $P_{n-k+1}$ and $P_{n-k}$, respectively, in units of 
the pixel size $\Delta$:
\begin{equation}
\label{a9}
I_n(\mu) = \Bigl(\mu a_n - a_{n-k}\Bigr) I_{n-k+1} + 
\Bigr(\mu(1-a_n) + a_{n-k}\Bigl) I_{n-k} \,,\quad
\mu_{nk}^+\leq \mu \leq \mu_{nk}^-\,,
\end{equation}
where $a_n= n+1/2$. The integral over the interval $[\mu_a,1]$
in (\ref{a1}) is split into the sum of integrals 
over the intervals in which the approximations (\ref{a8}) and
(\ref{a9}) hold
\begin{equation}
\label{a10}
I_n^\Omega = \sum_{k=1}^{N_n}\left( 
\int_{\mu_{nk}^+}^{\mu_{nk}^-} +
\int_{\mu_{nk+1}^-}^{\mu_{nk}^+}
\right) 
\frac{d\mu\, \mu}{\sqrt{1-\mu^2}}\ I_n(\mu)\,.
\end{equation}
The integrals are easily evaluated and the desired system 
of linear equations for $I_n$ is obtained. There is a simple
recurrence algorithm to solve it. Before describing it,
the issue about the integration limits in the term $k=N_n$
must be addressed.

Consider the last three integrals in the sum (\ref{a10}):
\begin{equation}
\label{a11}
\int_{\mu_{nN_n-1}^-}^{\mu_{nN_n}^+} + 
\int_{\mu_{nN_n}^+}^{\mu_{nN_n}^-} + 
\int^{\mu_{nN_n}^+}_{\mu_a}\,,
\end{equation}
in accord with the definition (\ref{a6}). 
By examining the possible overlaps of $P_n(\mu)$ with 
$P_{n-N_n}$ and $P_{n-N_n+1}$ depending on the value of
$\mu_a$, the following 
three possibilities are found to occur.
If $\mu_a\geq \mu_{nN_n}^-$ where the 
value of $\mu_{nN_n}^-$ is computed by the rule (\ref{a6}),
then the last two integrals in (\ref{a11}) cannot occur
in the sum (\ref{a10}) and must be removed. If 
$\mu_{nN_n}^+\leq \mu_a< \mu_{nN_n}^-$ where the values of
$\mu_{nN_n}^\pm$ are calculated by the rule (\ref{a6}),
then the very last integral in (\ref{a11}) cannot be present
and must be removed
from the sum (\ref{a10}). Only if $\mu_a <\mu_{nN_n}$ all three
integrals (\ref{a11}) are present in the sum (\ref{a11}).
For algorithm implementation purposes, it is convenient to 
retain the sum (\ref{a10}) as it is. The above three cases
can be accounted for if, {\it after} calculating 
the values of $\mu_{nk}^\pm$ by the rule (\ref{a6}), the values
of $\mu_{nN_n}^\pm$ are {\it redefined} by the rule:
$$
\mu_{nN_n}^- \rightarrow \max(\mu_a, \mu_{nN_n}^-)\,,\quad
 \mu_{nN_n}^+ \rightarrow \max(\mu_a, \mu_{nN_n}^+)\,.
$$
It is straightforward to verify that the unwanted integrals
in (\ref{a11}) (and, hence, in (\ref{a10})) always vanish
as their upper and lower limits coincide.

Define the functions
\begin{eqnarray*}
F_0(\alpha,\beta) &=& \int_\alpha^\beta \frac{d\mu\,\mu}
{\sqrt{1-\mu^2}} = \sqrt{1-\alpha^2} -\sqrt{1-\beta^2}\,,\\
F_1(\alpha,\beta)&=&\int_\alpha^\beta \frac{d\mu\,\mu^2}
{\sqrt{1-\mu^2}} = \frac 12\Bigl(\cos^{-1}(\alpha) -\cos^{-1}(\beta)
\Bigr) +\frac 12\Bigl(
\alpha\sqrt{1-\alpha^2} -\beta\sqrt{1-\beta^2}\Bigr)\,.
\end{eqnarray*}
They are used to evaluate the integrals in the sum (\ref{a10}).
For the central pixel $(n=0)$, one finds
\begin{equation}
\label{a12} 
I_0 = \frac{I_0^\Omega}{F_1(\mu_a,1)}\,.
\end{equation}
For $n=1,2,...,N$, Eq. (\ref{a10}) yields the recurrence relation
\begin{equation}
\label{a13}
I_{n} = \frac{I_n^\Omega - I_n^F}{C_{n1}}\,,
\end{equation}
where
\begin{eqnarray*}
C_{nk} &=& F_1(\mu_{nk}^+,\mu_{nk}^-)a_n - F_0(\mu_{nk}^+,\mu_{nk}^-)
a_{n-k}\,,\quad k=1,2,...,N_n\,,\\
I_n^F &=& \sum_{k=2}^{N_n} C_{nk}I_{n-k+1} +
\sum_{k=1}^{N_n}\Bigl[F_1(\mu_{nk+1}^-,\mu_{nk}^-) -
C_{nk}\Bigr] I_{n-k}\,.
\end{eqnarray*}
Note that $I_n^F$ depends only on $I_0$, $I_1$,...,$I_{n-1}$.
So by executing the recurrence relation (\ref{a13}) with the 
initial condition (\ref{a12}), all the values $I_n$ are found.
They can then be used to determine the emissivity by a suitable
Abel inversion algorithm. There is a vast literature on the numerical
Abel inversion (see, e.g., \cite{abelinversion}), 
and, for this reason, it will not be discussed here.
In Section 7, it is demonstrated that for emissivity functions
representative for LIBS plasmas, the proposed algorithm provides
a very accurate reconstruction of $I_n$. Thus, the accuracy
in the emissivity reconstruction would be determined by the accuracy
of a particular Abel inversion algorithm.

\subsubsection{General apertures}  

It is not difficult to extend the proposed algorithm 
to an aperture of a generic shape. Suppose that 
$A$ is symmetric under the reflection $u\rightarrow -u$
and $A_+$ is the half of $A$ for which $u\geq 0$.
A generalization to the non-symmetric case is trivial.
Integrating Eq. (\ref{4}) over the interval occupied
by the $n$th pixel, a generalization of (\ref{a1}) is 
obtained
\begin{equation}
\label{g1}
I_n^\Omega = 2\int_{A_+} dA\, \frac{f^2}{(f^2+u^2+v^2)^2}\ 
I_n(\mu(u,v))\,,
\end{equation}
where the function $\mu(u,v)$ is defined by the second equation
in (\ref{4}). The area $A_+$ can be partitioned into regions 
that lies between the level curves (hyperbolas) 
of the function $\mu(u,v)$:
$$
\mu(u,v) =\mu_{nk}^+\,, \quad \quad \mu(u,v)=\mu_{nk}^-\,,
$$
where $\mu_{nk}^\pm$ are defined in (\ref{a6}) and 
$\mu_a = \min_{A_+}\mu(u,v)$. The procedure is shown in 
the right panel of Fig.~\ref{fig2}.
Let $A_{nk}^-$ and $A_{nk}^+$ 
be the regions
in which $\mu_{nk}^+\leq \mu(u,v)\leq \mu_{nk}^-$ and 
$\mu_{nk+1}^-\leq \mu(u,v)\leq \mu_{nk}^+$, respectively.
Then a generalization of Eq. (\ref{a10}) to the case of a 
generic aperture reads
\begin{equation}
\label{g2}
I_n^\Omega = 2\sum_{k=1}^{N_n}\left( 
\int_{A_{nk}^-} +
\int_{A_{nk}^+}
\right) dA\
\frac{f^2}{(f^2+u^2+v^2)^2}\ I_n(\mu(u,v))\,.
\end{equation}
The approximations (\ref{a8}) and (\ref{a9}) are used 
for $I_n(\mu)$ in the regions $A_{nk}^+$ and $A_{nk}^-$,
respectively to evaluate the double integrals in (\ref{g2}).
Depending of the shape of $A$, it can be done either analytically
or numerically. Put
$$
F_m(B) = 2 \int_B dA \ \frac{f^2}{(f^2+u^2+v^2)^2}\ \mu^m(u,v)\,,\quad
\quad m=0,1\,,
$$
for a planar region $B$.
The initial condition (\ref{a12}) and the
recurrence relation (\ref{a13}) become
\begin{equation}
\label{g3}
I_0 = \frac{I_0^\Omega}{F_1(A_+)}\,,\quad
I_n= \frac{I_n^\Omega - I_n^F}{C_{n1}}\,,\quad n=1,2,...,N\,,
\end{equation}
where
\begin{eqnarray*}
C_{nk}&=& a_n F_1(A_{nk}^-) -a_{n-k}F_0(A_{nk}^-)\,,\quad
a_n=n+1/2\,,\\
I_n^F &=& \sum_{k=2}^{N_n} C_{nk}I_{n-k+1} + 
\sum_{k=1}^{N_n}\Bigl(F_1(A_{nk})-C_{nk}\Bigr) I_{n-k}\,,
\end{eqnarray*}
and $A_{nk}$ is the union of $A_{nk}^+$ and $A_{nk}^-$.

\subsubsection{Numerical results for a hyperbolic aperture}

\begin{figure}
 %\centering
\includegraphics[height=4cm,width=17cm]{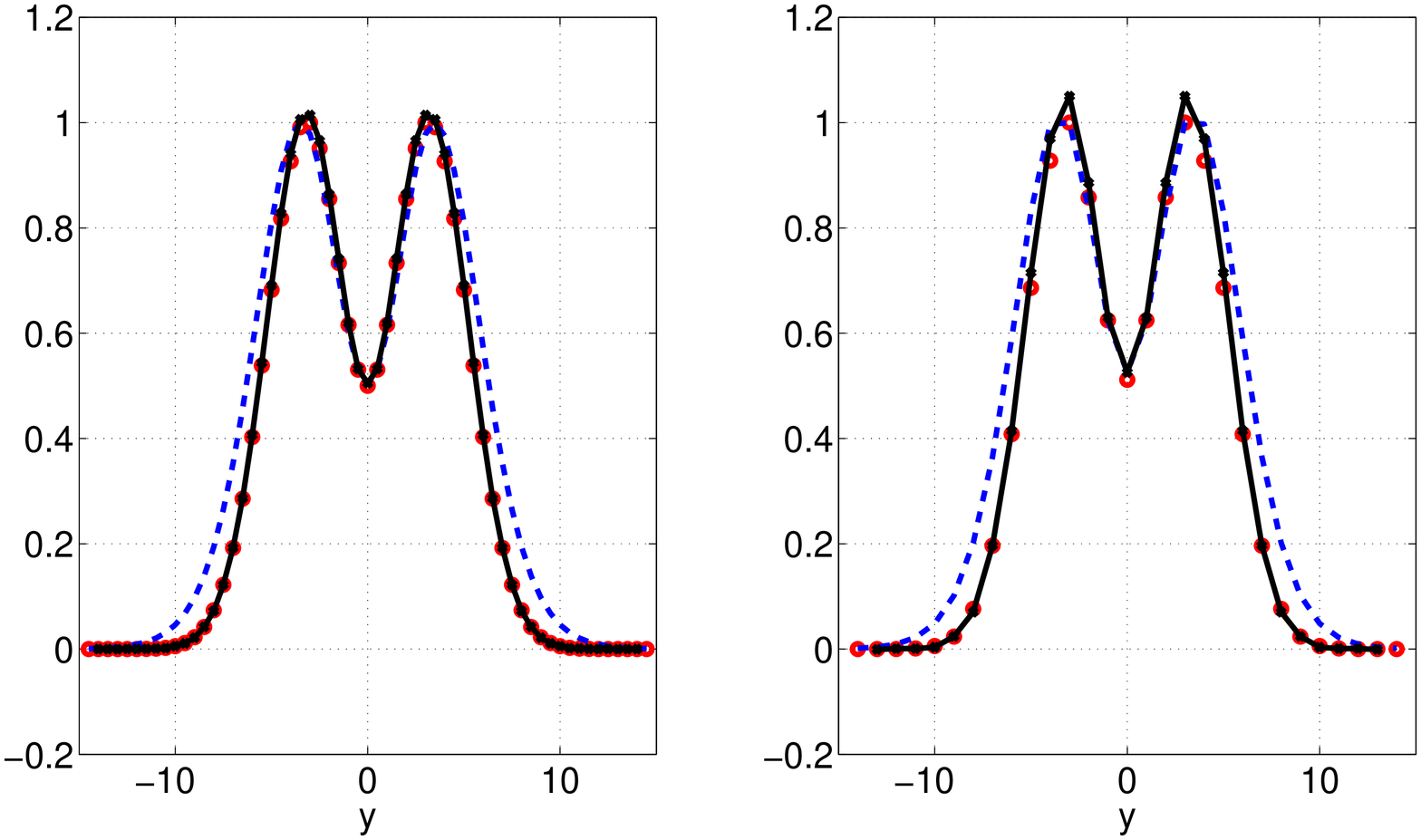}
\caption{\small Numerical results for a hyperbolic aperture
with $\mu_a=\cos\theta_a$, $\theta_a = 51.3^o$.
{\sf Left panel:} The solid (black) curve is the graph of 
the intensity $I(y)$ in parallel lines of sight for the 
emissivity $\ve(r)$ described in the text.
The dashed (blue) curve is the observed intensity $I_\Omega(y)$
calculated from $I(y)$ as specified in Eq. (\ref{7}). Both the 
curves are normalized on their maximal values. The amplification
of the signal is not shown explicitly. It is determined as
explained in the text.
Solid (red) dots
are reconstructed values of $I_n$ from $I_n^\Omega$. The number
of pixels is 60 ($N=31$).
{\sf Right panel:} The same as in the left panel but the 
number of pixels is 30 $(N=16)$. 
A small inaccuracy of 
the reconstructed values of $I_n$ is observed
near the maxima of the intensity $I(y)$ which occurs
through errors in the approximations (\ref{a8}) and (\ref{a9}).  
}
\label{fig3}
\end{figure}
The reconstruction algorithm has been tested with synthetic
data. The emissivity function is taken in the form
$\ve(r) = \ve_0 \exp(-r^2/(2\sigma^2))(a - \cos(br))$
where the values of the parameters (arbitrary units) 
are $\ve_0=1$, $\sigma =3$, $a=1.4$, and $b=0.5$. The 
emissivity profile is typical for the LIBS plasmas 
observed in experiments \cite{7}. The intensity $I(y)$ in parallel
lines of sight is calculated by means of Eq. (\ref{5}).
Then the integral intensity over all lines of sight 
within the acceptance solid angle defined by the hyperbolic
aperture with $\mu_a=\cos\theta_a$, $\theta_a=51.3^o$
is found by Eq. (\ref{8}) (the amplification factor
$4\sin\varphi_a$ is omitted as it is irrelevant for 
the reconstruction algorithm). The synthetic data are 
calculated by means of Eq. (\ref{a1}) when 60 and 30 pixels
are placed across the plasma plume diameter. Then the reconstruction 
algorithm is applied to find the values $I_n$. The results
are shown in Fig.~\ref{fig3}. The left and right panels represent
the cases with $60$ and $30$ pixels, respectively. 
The solid (black) and dashed (blue) curves are the graphs 
of $I(y)$ and $I_\Omega(y)$, respectively. 
Both the curves are normalized to their maximal values so that
the amplification factor due to the vertical size $\Delta_a$ 
of the hyperbolic aperture cannot be seen.
This normalization would not make sense for a generic aperture
because there is no factorization in the sum over angled lines
of sight for apertures of a general shape (see Section 6).
Yet, the factorization due to the hyperbolic aperture 
leads to a mild deviation of $I_\Omega(y)$ from $I(y)$ when they are
normalized on their maximal values. 
The solid (red) dots
show the reconstructed values of $I_n$. The reconstruction 
is very accurate for the case of $60$ pixels. When the pixel
size increases, the approximations (\ref{a8}) and (\ref{a9})
becomes less accurate as proved in Appendix. A reconstruction
error increases near the maxima of $I(y)$
when the pixel size is doubled (the right
panel of Fig.~\ref{fig3}). For a typical pixel detector,
the pixel size is about $0.02\, mm$. So, for a typical 
LIBS plasma plume diameter of $2\, mm$, there are about 
100 pixel collecting the data.

\renewcommand\thesubsubsection{A.}
\subsubsection{Appendix}
Here the accuracy of the approximation (\ref{a7}) is assessed.
For a continuous function $f(x)$ there exist points
$a\leq x_*\leq b$ and $a'\leq x'_*\leq b'$ such that 
$$
I=\int_a^b dx f(x) = (b-a) f(x_*)\,,\quad
I'=\int_{a'}^{b'} dx f(x) = (b'-a')f(x_*')\,.
$$
It  follows from these relations that
$$
I' = \frac{\Delta'}{\Delta}\, I +\Delta' \Bigl(f(x_*')-f(x_*)\Bigr)\,,
$$
where $\Delta = b-a$ and $\Delta'=b'-a'$.
By the mean value theorem, the last term in this equality 
can be transformed so that
$$
I' = \frac{\Delta'}{\Delta}\, I +\Delta' (x_*'-x_*)f'(\xi)\equiv
\frac{\Delta'}{\Delta}\, I + \gamma
$$
for some $\xi$ between $x_*'$ and $x_*$. Therefore the error 
$\gamma$ is bounded by
$$
|\gamma| \leq M \Delta'\Delta\,,\quad \quad M=\max_{[a,b]}|f'(x)|\,.
$$
If $\Delta\rightarrow 0$, while $\Delta'/\Delta$ remains finite,
the error decreases as $\Delta^2$.

\renewcommand\thesubsubsection{Acknowledgments}

\subsubsection{}

The authors acknowledge stimulating discussions with 
Prof. U. Panne (BAM) and D. Shelby (UF, Chemistry).
S.V.S. is grateful to Prof. U. Panne for
his continued support and thanks Department IV of BAM for a kind 
hospitality extended to him during his visit.
The work of I.B.G. is supported in part by 
the DFG-NSF grant GO 1848/1-1 (Germany) and 
NI 185/38-1 (USA).

\end{document}